\documentclass[utf8]{frontiers-astro} % for Science, Engineering and Humanities and Social Sciences articles
\def\keyFont{\fontsize{8}{11}\helveticabold }
\def\firstAuthorLast{F. Combes} %use et al only if is more than 1 author
\def\Authors{F. Combes\,$^{1,*}$}
\def\Address{$^{1}$Observatoire de Paris, LERMA, College de France, CNRS, PSL, Sorbonne University UPMC, Paris, France}
\def\corrAuthor{F. Combes}

\def\corrEmail{francoise.combes@obspm.fr}

\begin{document}
\onecolumn
\firstpage{1}

\title[AGN feedback]{AGN feedback and its quenching efficiency} 

\author[\firstAuthorLast ]{\Authors} %This field will be automatically populated
\address{} %This field will be automatically populated
\correspondance{} %This field will be automatically populated

\extraAuth{}% If there are more than 1 corresponding author, comment this line and uncomment the next one.
%\extraAuth{corresponding Author2 \\ Laboratory X2, Institute X2, Department X2, Organization X2, Street X2, City X2 , State XX2 (only USA, Canada and Australia), Zip Code2, X2 Country X2, email2@uni2.edu}

\maketitle

\begin{abstract}
In the last decade, observations have accumulated on gas outflows in 
galaxies, and in particular
massive molecular ones. The mass outflow rate is estimated
between 1-5 times the star formation rate. For the highest
maximal velocities, they are driven by AGN; these
outflows are therefore a clear way to moderate or suppress star formation.
Some of the most convincing examples at low redshift come from the radio mode,
when the radio jets are inclined towards the galaxy plane, or expand
in the hot intra-cluster medium, in cool core clusters.  
However, AGN feedback can also be positive in many occasions, and
the net effect is difficult to evaluate. The quenching
efficiency is discussed in view of recent observations.

\section

\tiny
 \keyFont{ \section{Keywords:} galaxies, active galaxy nuclei, black holes, outflows, molecules} %All article types: you may provide up to 8 keywords; at least 5 are mandatory.
\end{abstract}

\section{Types of feedback}
Cosmological simulations in the CDM scenario predict too many galaxies at both ends of the mass function. If it is possible to suppress star formation through supernovae feedback in dwarf galaxies, we have to rely on AGN feedback to quench star formation in massive galaxies.
There are two main modes of AGN feedback: first, the
quasar mode, called also radiative mode or wind mode.
This occurs when the AGN luminosity is high, 
close to Eddington, mainly for young QSO at high redshift.
Due to radiation pressure on the ionized gas, the nucleus
reaches its gas accretion limit, and begins to eject some gas in a wind. 
Since the Eddington luminosity $L_{Edd}$  is proportional
to $M_{BH}/\sigma_T$, where  $M_{BH}$ is the supermassive black hole mass and
 $\sigma_T$  the Thomson
cross section, the Eddington limitation in BH growth might
explain the M-$\sigma$ relation, i.e.  $M_{BH}  \propto f \sigma_T \sigma^4$,
where  $f$ is the gas fraction, and $\sigma$ the central velocity dispersion.
The same consideration can be made, when a central starburst
reaches its Eddington luminosity, with radiation pressure on dust. Now the cross
section is $\sigma_d$, which is 1000 times higher than  $\sigma_T$.
This could lead to a limitation of the bulge mass to 1000 $M_{BH}$,
quite close to the observed $M_{bulge}/M_{BH}$ ratio (Fabian 2012).

The second feedback scenario is the radio mode, or kinetic mode, due to radio jets.
This takes place in very low luminosity AGN,
when L $<$ 0.01 $L_{Edd}$, mainly at low redshift. It is frequent in relatively massive galaxies, 
like the radio-loud ellipticals, powered by a
radiatively inefficient flow (ADAF). A particular example of
this feedback mode is the moderation of  
cooling flows in clusters, through radio-jets from the brightest central galaxy (BCG).
It is observed also in low-luminosity AGN, like Seyfert galaxies (Combes et al. 2013, Dasyra et al. 2016).

\begin{figure}[h!]
\begin{center}
\includegraphics[width=10cm]{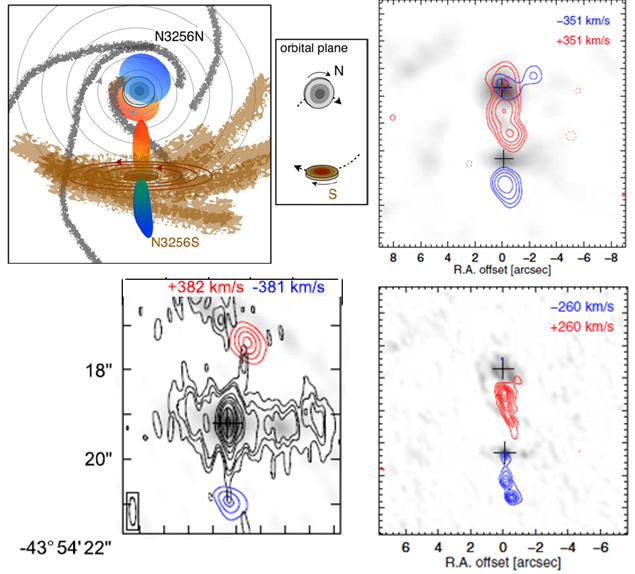}
\end{center}
%\caption{ Repeat as  necessary for each of your figures}
{{\bf Figure 1.} ALMA observations of two molecular outflows in the merging galaxies NGC3256 (sketched at top left), 
adapted from Sakamoto et al. (2014), figure reproduced with permission.
The northern object, almost face-on, reveals an outflow towards the observer, while the southern
galaxy, nearly edge-on, shows a collimated outflow in the plane of the sky. The latter appears
to coincide with a small radio jet, according to the VLA 3.6cm radio map at the bottom left.
The two CO maps at the right (top at low resolution, and bottom at high resolution) show that the northern
outflow is resolved out. It must be a wide-angle outflow, uncollimated.}
\label{fig1}
\end{figure}

Frequently, star formation and nuclear activity are associated, and it is difficult
to disentangle the supernovae and AGN feedback.  Galactic winds coming from 
a starburst (like the prototypical M82) are in general less violent, with smaller 
maximum velocity, and un-collimated. An example of both is provided
by the galaxy merger NGC 3256, an ultra-luminous starburst at z=0.01.
ALMA observations of the molecular gas (through the CO(3-2) line) have revealed
high-velocity wings in both nuclei, the face-on N3256N and almost edge-on N3256S
(Sakamoto et al 2014).
For the latter, the outflow is highly collimated, and likely due to an AGN (cf Figure \ref{fig1}).
The derived maximum velocity is $\sim$2000 km/s out to 300 pc, and corresponds
to 50 M$_\odot$/yr. For the northern galaxy, the maximum velocity is $\sim$750 km/s,
and the outflow rate of  60 M$_\odot$/yr. In both cases, these rates are comparable to the
star formation rate, showing that the implied quenching is significant. 
The time-scale to develop these outflows is $\sim$ 1 Myr (Sakamoto et al 2014).

Numerical simulations have begun to study the
radiative mode, depending however on small-scale recipes,
which are calibrated on observations. Recipes are required to
take into account the black hole growth and its associated 
feedback, all being sub-grid physics. Different groups
do not converge to the same conclusions.
While Springel et al. (2005) and Hopkins et al. (2006) found a good coupling
between the AGN and the galaxy, Gabor \& Bournaud (2014) conclude that
the quasar mode has no quenching effect. Observations are therefore
key to solve the issue of the AGN quenching efficiency.

\section{Cool-core clusters}

One example where AGN feedback is clearly demonstrated 
is found in the center of cool-core clusters. It has been known for a long
time that the cooling time-scale of the hot ICM gas becomes smaller
than the Hubble time in the center, and cooling flows are expected. However,
only 10\% of the expected cooling rate is observed, and this is now
understood to be due to the radio jets of the central AGN reheating the 
gas. The jets carve cavities in the ICM, and uplift some hot gas. The 
denser regions around cavities cool in filaments, which infall after losing
their pressure support, and are conspicuous
in H$\alpha$ (shocks) and  molecular gas (Salom\'e et al. 2006, 2008).

Although most H$\alpha$ is excited by shocks, there are 
some clumps of star formation (Canning et al. 2014). The observed scenario
is far from the simple model of cooling flows, where most of the cooling occurred
in the center. In real clusters, gas cools in the border of cavities,
which can occur 20-50kpc from the center. When the central galaxies is
not at rest but oscillating in the cluster potential well, a cooling wake 
extends over $\sim$ 50 kpc (Salom\'e \& Combes 2004, Russell et al. 2017).
In this complex picture, gas inflow and outflow coexist; 
the molecular gas coming from previous cooling is dragged out by 
the AGN feedback, and can explain the large metallicity, necessary
to detect the CO lines.
The uplifted bubbles of hot gas create inhomogeneities and further cooling,
and the cooled gas fuels the AGN, to close the loop. Through ram-pressure forces,
the cold gas velocity is much lower than free-fall (Salom\'e et al. 2008).

\section{Molecular outflows}

Molecular outflows are now frequently observed in nearby
galaxies, and statistics have been made with respect
to their starburst or AGN origin (Cicone et al. 2014).
For AGN-host galaxies, the outflow rate correlates with the AGN power,
and also the L$_{AGN}$/L$_{bol}$ luminosity ratio (L$_{bol}$ being the total luminosity of the galaxy, including the AGN). The correlation
does not exist for starbursts.
What is also highly interesting is the good correlation between the momentum
carried in the outflow (vdM/dt, where v is the outflow velocity )with the photon momentum output of the AGN 
L$_{AGN}$/c.  The average value is vdM/dt  $\sim$ 20 L$_{AGN}$/c, which
is only possible with energy-driven outflows (Zubovas \& King 2012), i.e. when the energy injected by the inner wind is fully conserved throughout the outflow, unless the outflowing shell is optically thick to the infrared radiation, implying high momentum flux (Ishibashi \& Fabian 2015).

For the radio mode to be efficient in quenching star formation, there must be
a strong coupling with galaxy disks. This is the case when the radio jet is 
not perpendicular to the galaxy plane, but is inclined so that the jet can sweep
out some significant region in the disk.  For example, the   
radio jet starts its way in the plane of the Seyfert 2 galaxy NGC1068:
a molecular outflow of 63 M$_\odot$/yr, about 10 times the SFR has been
observed by ALMA in the circum nuclear disk region 
(Garcia-Burillo et al. 2014). In the extreme case of IC5063, 
the jet is entirely in the plane, and creates secondary outflow
features, at each collision with in-plane clouds
(Morganti et al. 2015, Dasyra et al. 2016).
Some of the gas might be optically thin in theses flows.

\begin{figure}[h!]
\begin{center}
\includegraphics[width=12cm]{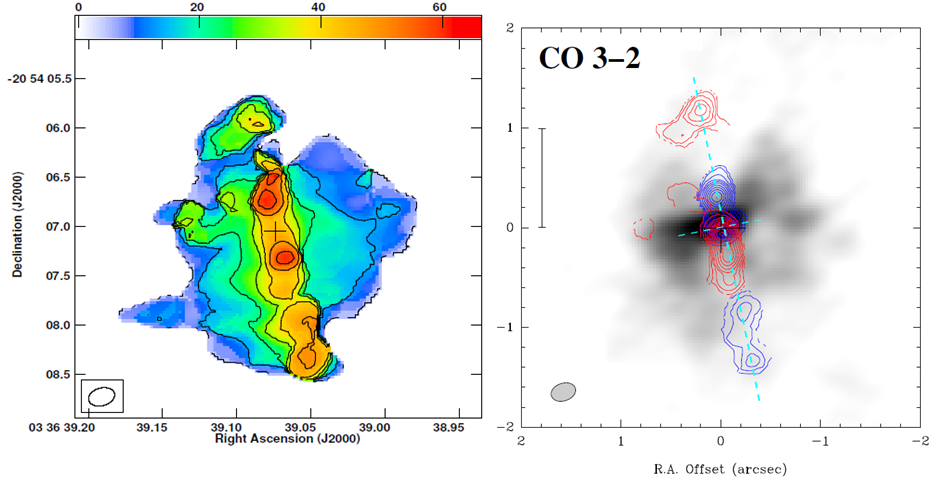}
\end{center}
%\caption{(A) is one logo, (B) is a different logo.}
{{\bf Figure 2.} The highly collimated molecular outflow in NGC1377, from Aalto et al. (2016), 
figure reproduced with permission.
The left map is the velocity dispersion, and the right shows the blue and red-shifted CO emission. Note that the
velocity of the flow changes sign at mid-way, symmetrically in the North and South.}
\label{fig2}
\end{figure}

Feedback is also observed in low-luminosity AGN.
One of the smallest outflow detected up to now is
that of the Seyfert 2 NGC1433, with a maximum outflow
of 100 km/s, along the minor axis (Combes et al. 2013).
The case of the lenticular NGC1377 is very puzzling (Aalto et al. 2016).
There is no radio AGN emission, although a very collimated molecular outflow
is detected, with even a precession, visible since the outflow is almost in the
plane of the sky (see Figure 2).

\begin{figure}[h!]
\begin{center}
\includegraphics[width=14cm]{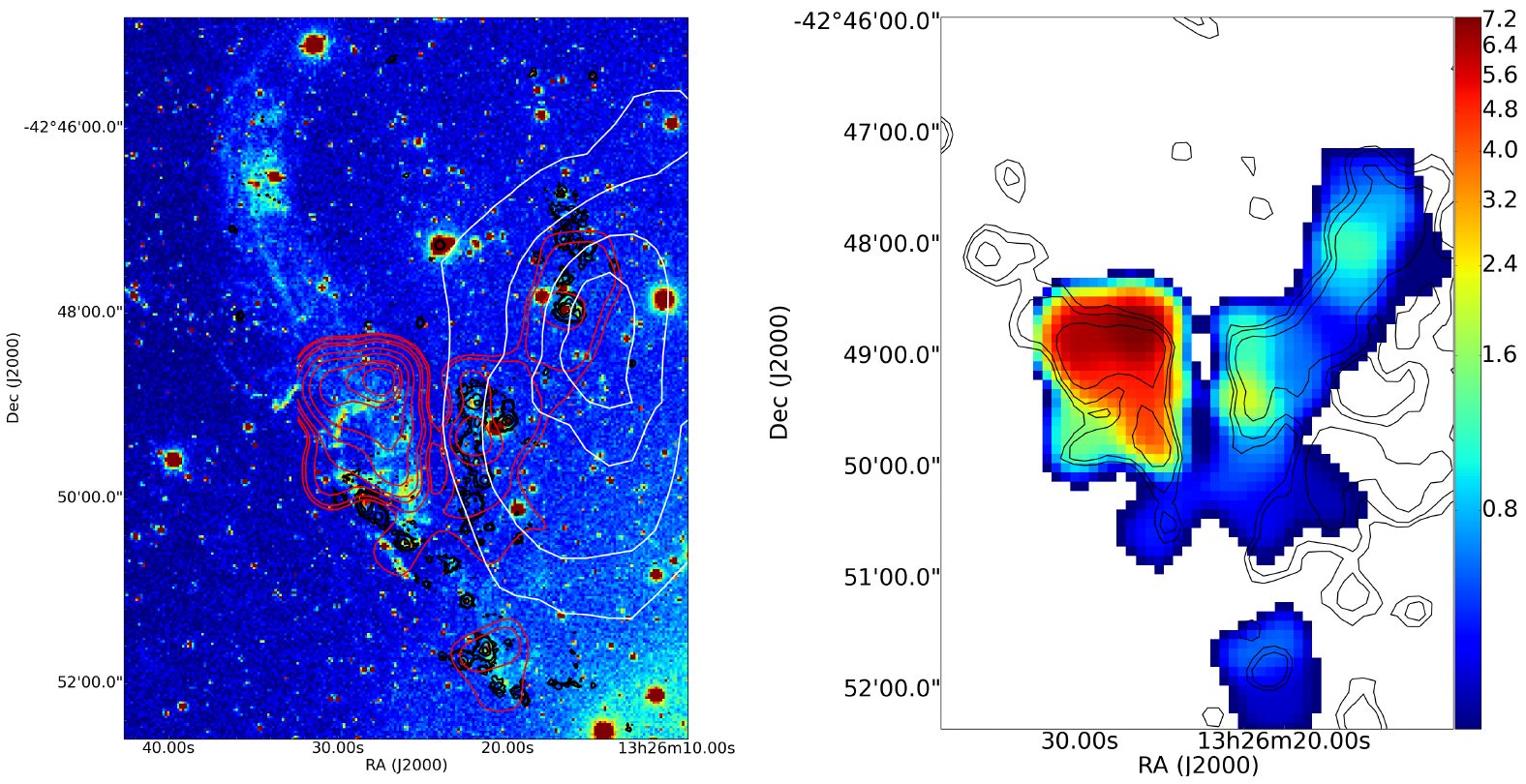}
\end{center}
{{\bf Figure 3.} Multi-phase filament in the radio jet of Centaurus A, from Salom\'e et al. (2016b),
figure reproduced with permission.
{\bf Left}: Superposed on the H$\alpha$ map, the red contours correspond to CO emission, 
the white to HI, and the black to FUV from GALEX. {\bf Right}: the black HI contours are superposed on
the CO emission map, showing that most of the molecular gas is outside the HI region.} 
\label{fig3}
\end{figure}

\section{Jet-induced star formation}

The AGN feedback is frequently negative, but can be also positive,
and trigger star formation. 
One of the most convincing examples of jet-induced star formation
has been found in the Minkowski object (NGC541, distance of 82 Mpc),
where conspicuous HII regions are observed at the extremity of the
radio jet, outside of the optical galaxy  (Croft et al. 2006).
Molecular gas has been found in NGC541, but only an upper limit
in the HII region with IRAM (Salom\'e, Salom\'e, Combes 2015),
suggesting a high efficiency of star formation.
Recently, CO emission was  detected with ALMA (Lacy et al. 2017),
compatible with the IRAM upper limit. However, when
taken into account the low metallicity of the gas, which increases the  CO-to-H$_2$ conversion
ratio and thus the gas mass, and the excitation by shocks (reducing the SFR, for a given H$\alpha$ flux), the star formation efficiency 
is now much lower in the triggered region.

A more nearby example of jet-induced star formation has been studied
in Centaurus A, at a distance of 3.4 Mpc. Atomic gas has been mapped in shells around the galaxy
by Schiminovich et al. (1994), and molecular gas has been found 
in those shells on the path of the radio jet by Charmandaris et al. (2000).
Near the northern shell, there is a conspicuous filament of star formation,
aligned along the jet, mapped in H$\alpha$, FUV-GALEX, and dust
emission with Herschel. Molecular gas has been mapped with APEX
and ALMA, and suprisingly, there is even more CO detected outside the
HI shell that inside (Salom\'e et al. 2016b).
The molecular gas is concentrated at the region where the jet encounters
the HI shell, and the atomic gas is then transformed in the molecular phase
under the impact of the jet pressure (see Figure \ref{fig3}). 
 Ionised gas excitation was determined with MUSE spectral observations,
with the help of the BPT diagnostics (Salom\'e et al 2016a), and is
mainly due to shocks, with some contribution of star formation.

There is clearly star formation triggering from the
radio jet, however, the star formation efficiency
is lower than in galaxy disks. This is a situation comparable
to what is found in the outer parts of galaxies, where gas layers
are flaring (e.g. Dessauges-Zavadsky et al. 2014).
The reason might be a lack of pressure, due to low or absent
restoring force from a stellar disk.
The importance of pressure, and of the surface density of stars
for the star formation  efficiency has been emphasized by 
Blitz \& Rosolowsky (2006), and Shi, Helou et al. (2011).

\section{Conclusions}

AGN feedback is required to quench star formation in massive
galaxies, to reproduce the observed galaxy mass function, and
avoid the over-production of very massive galaxies in cosmological
simulations. One can consider two types 
of AGN feedback: the quasar mode more frequent
at high redshift, and the radio mode, more easy to observe
in nearby galaxies.

One environment where the AGN feedback efficiency is clear
is represented by cool core clusters, where the radio jets of
the central bright galaxy carve bubbles and cavities in the hot
intra-cluster gas, and moderate the gas cooling.
Nearby galaxies frequently reveal significant molecular outflows,
with a loading factor, the ratio between the outflow rate and the star
formation rate, between 1 and 10. Given the high momentum rate, the
outflows appear to be energy conserving.

AGN feedback can also be positive. Some evidence of jet-induced star formation has been observed. In particular, the high jet
pressure can trigger the phase transformation from atomic to molecular
gas, favoring star formation. This triggered star formation
is however less efficient than in normal galaxy disks.

\section*{Acknowledgments}
All appropriate permissions have been obtained from the copyright holders 
of the figures reproduced in the manuscript.
Many thanks to Mauro d'Onofrio and the organising committee for this exciting conference on Quasars
in Padova, on April 2017.

\section*{References}
%\begin{thebibliography}
 Aalto, S., Costagliola, F., Muller, S. et al.: 2016, A\&A 590, A73 % N1377
\\ Blitz, L., Rosolowsky, E.: 2006, ApJ  650, 933
\\ Canning, R. E. A., Ryon, J. E., Gallagher, J. S. et al.: 2014, MNRAS  444, 336
\\ Charmandaris, V., Combes, F., van der Hulst, J. M.: 2000, A\&A  356, L1
\\ Cicone, C., Maiolino, R., Sturm, E. et al.: 2014, A\&A   562, A21
\\ Combes, F., Garcia-Burillo, S., Casasola, V. et al.: 2013 A\&A, 558, A124 %N1433
\\ Croft, S., van Breugel, W., de Vries, W. et al.: 2006, ApJ  647, 1040
\\ Dasyra, K. M., Combes, F., Oosterloo, T. et al.: 2016 A\&A 595, L7
\\ Dessauges-Zavadsky, M., Verdugo, C., Combes, F., Pfenniger, D.: 2014, A\&A 566, A147 
\\ Fabian, A. C.: 2012, ARA\&A, 50, 455
\\ Gabor, J. M., Bournaud, F.: 2014, MNRAS  441, 1615
\\ Garcia-Burillo, S., Combes, F., Usero, A.:  2014, A\&A 567, A125 % N1068
\\ Hopkins, P., Hernquist, L., Cox, T. J. et al.:  2006, ApJS 163, 1
\\ Ishibashi, W., Fabian, A. C.: 2015, MNRAS 451, 93
\\ Lacy, M., Croft, S., Fragile, C. et al.: 2017, ApJ 838, 146
\\ Morganti, R., Oosterloo, T., Oonk, J. B. R. et al.: 2015 A\&A   580, A1
\\ Russell, H.R., McNamara, B.R., Fabian, A.C. et al.: 2017, MNRAS, submitted 
\\ Sakamoto, K., Aalto, S., Combes, F. et al.: 2014, ApJ 797, 90
\\ Salom\'e P., Combes F., 2004, A\&A, 415, L1
\\ Salom\'e, P., Combes, F., Edge, A. C. et al.: 2006, A\&A 454, 437
\\ Salom\'e, P., Combes, F., Revaz, Y. et al.: 2008, A\&A 484, 317
\\ Salom\'e, Q., Salom\'e, P., Combes, F.: 2015, A\&A  574, A34 % MO
\\ Salom\'e, Q., Salom\'e, P., Combes, F., Hamer, S., Heywood, I: 2016a, A\&A  593, A45 %  SF efficiency CenA
\\ Salom\'e, Q., Salom\'e, P., Combes, F., Hamer, S.: 2016b, A\&A  595, A65 % CenA HI-H2
\\ Schiminovich, D., van Gorkom, J., van der Hulst, T., Kasow, S.: 1994, ApJ 423, L101
\\ Shi, Y., Helou, G., Yan, L. et al.: 2011, ApJ  733, 87
\\ Springel, V., Di Matteo, T., Hernquist, L.: 2005, MNRAS  361, 776
\\ Zubovas, K., King, A.: 2012, ApJ  745, L34
%\end{thebibliography}
%%% Make sure to upload the bib file along with the tex file and PDF
%%% Please see the test.bib file for some examples of references

\end{document}